\documentclass{IAU-TrB}
\usepackage{graphicx}
\newcommand{\NOPRINT}[1]{\null}

\title[ASTROMETRY]     %% header right hand page %%
{}

\author[DIVISION I COMMISSION 8]   %% header left hand page %%
{}

\pubyear{2015}
\volume{Volume XXIXA}
\pagerange{1--4}
\date{15 Oct 2015}
\jname{Transactions IAU, Volume XXIXA}
\editors{Thierry Montmerle, ed.}
\begin{document}

\maketitle

{\bf

\large
\begin{tabbing}
\hspace*{65mm}       \=                                     \kill
DIVISION I \\COMMISSION 8         \> ASTROMETRY              \\
                     \> {\it (ASTROMETRY)}                   \\
\end{tabbing}

\normalsize

\begin{tabbing}
\hspace*{65mm}       \=                                     \kill
PRESIDENT            \> Norbert Zacharias  \\
VICE-PRESIDENT       \> Anthony Brown      \\
PAST PRESIDENT       \> Dafydd Evans       \\
ORGANIZING COMMITTEE \> Li Chen, Naoteru Gouda, \\
                     \> Valeri Makarov, Aleksandr Shulga, \\
                     \> Jean Souchay, Rama Teixeira,\\ 
                     \> Stephen Unwin \\
\end{tabbing}

% \bigskip
% \noindent
% REPORT 2012 -- 2015
}

%\small


\vspace*{-10mm}
%\vspace{3ex}

\section{Introduction}

\noindent 
Commission 8 has regularly published triennial reports in the past
and the current OC therefore voted to adopt a traditional format also
for this special Legacy issue of the IAU Transactions.
The outgoing President is grateful for the support of many Commission 
members who contributed to this report.  Our contribution consists of 
3 parts: 1) this introduction, providing a general overview and 
highlights of recent research in astrometry, 2) a summary of the 
astrometry business \& science meeting at the 2015 IAU General Assembly, 
and 3) the activity report of our Commisson covering the mid-2012 to 
mid-2015 period.

Astrometry is about to be revolutionized by the European Space
Agency (ESA) Gaia mission.  Regular survey operations began mid-2014.
The first data release is expected by mid-2016 and the final catalog 
by about 2020, depending on the lifetime of the mission, currently 
expected to be over 5 years.  Gaia will provide an optical reference 
frame with positions, proper motions and parallaxes in the 10 to 300 
micro-arcsecond ($\mu$as) range (depending on brightness) down to
almost 21st magnitude, which will soon make most current
optical reference star catalogs obsolete.

Radio astrometry continues to provide the defining celestial
reference frame by VLBI observations of compact, extragalactic
sources.  The 2nd version of the International Celesctial
Reference Frame (ICRF2) was adopted by the IAU in 2009,
and observations and reductions are in progress for the
ICRF3, scheduled to be adopted by the next IAU GA in 2018.
The link between the Gaia (optical) and the ICRF (radio)
coordinate systems is being investigated currently,
however only a small number of sources seem to be in common and 
optical structure of AGNs is posing a possible challenge.
ALMA in Chile now coming on-line and the SKA projects will be very 
valuable for astrometry in the coming decade.

Another highlight in recent astrometry is the ability to obtain 
micro-arcsec parallaxes from radio interferometry (VLBI and the 
Japanese VERA project).
Of particular importance was the distance estimate of Pleiades 
stars (Melis et al. 2014) which agrees with earlier ground-based 
results and our current understanding of stellar evolution models, 
while suggesting that the Hipparcos parallax of the Pleiades is in 
error by more than previously expected.

The 4th and final US Naval Observatory (USNO) CCD Astrograph Catalog 
was published in 2012, providing a dense, accurate reference frame on
the ICRF at optical wavelenghts.  A deeper and even more accurate
star catalog, the USNO Robotic Astrometric Telescope (URAT) Catalog 
was published 
in 2015. This likely concludes the long history of such ground-based
endeavors to provide an optical reference frame in light of the
upcoming Gaia 1st data release.  The future large synoptic survey 
telescope (LSST) will utilize the Gaia reference frame and go much 
deeper than Gaia with high frequency of sky coverages and multiple
bandpasses.  PanSTARRS operated successfully throughout the last
couple of years as precursor to LSST, providing unique astrometric 
data for objects fainter than UCAC4 or URAT.

The long anticipated textbook, {\em Astrometry for Astrophysics}
(Ed. W.~van Altena) was published in 2013 by Cambridge University Press.
This is a community effort with many participating authors explaining
methods, models and applications of our science to the next generation
of astronomers.

\section{Summary of the Commission 8 meeting at the 2015 GA}

A total of 4 hours were graciously allocated for the Commission 8
meeting, that included several science presentations preceded by
a traditional business meeting of the Commission.  
This meeting was attended by about 30 people and all presentations
are available from the Commission's web sites at 
www.ast.cam.ac.uk/ioa/iau\_comm8/ and www.iau.org .

\subsection{Business meeting}

The outgoing President, Norbert Zacharias presented an overview
about the re-organization of the IAU Divisions and Commissions.
All current Commissions were dissolved and new Commissions established.
Following the recommendation of the Organizing Committee (OC) of the 
Astrometry Commission,
a proposal was assembled by Anthony Brown, Dafydd Evans and Norbert
Zacharias to establish a ``new" Commission with the same name and with 
the same goals as the old Commission.  This proposal was accepted by the 
IAU and assigned the designation ``CA.1" to indicate its parent Division A.

The incoming President is Anthony Brown (Netherlands) and the elected 
new Vice-President (VP) is Jean Souchay (France).  Alexandre Andrei (Brazil),
Yoshiyuki Yamada (Japan), and Stephen Unwin (USA) are the elected
new members of the OC.  The co-proposer of the new Astrometry Commission,
Dafydd Evans (UK) serves again on the OC as well as
Norbert Zacharias ex officio.

The Commission membership has increased to 283 as of July 20, 2015.
The following deceased members of our Commission were honored by
a moment of silence: Donald Backer (Univ. California, Berkeley),
Hans-Heinrich Bernstein (Heidelberg), Arkadiy Kharin (Main Astronomical
Observatory Ukraine), Evgenia Khrutskaya (Pulkovo Observatory),
Vasile Mioc (Astronomy Inst. Romanian Academy of Sciences),
Naufal Rizvanov (Kazan State University), and Haruo Yasuda (Tokyo).

During the past triennium a total of 10 newsletters were issued to all 
members of the Commission to inform about highlights and news in our
field of research, related meetings, and relevant IAU activities.
The following key projects and programs were mentioned:  the Gaia 
mission, work on the ICRF3, the URAT1 catalog, Standards Of Fundamental
Astronomy (SOFA), PanSTARRS, Skymapper, LSST, VERA, WG Astrometry
by Small Ground-based Telescopes, LQAC3, and SDSS DR12.
No questions or additions from the audience were received at the
conclusion of the business meeting.

\subsection{Science presentations}

Jean-Eudes Arlot, on behalf of Marcelo Assafin, reported on the
activities of the WG Astrometry by Small Ground-based Telescopes.
This WG is under Division A and its continuation is recommended to
coordinate research efforts and assist in astrometric education 
and training.

Jean-Eudes Arlot presented a talk about astrometry of solar system 
objects after Gaia.  He domonstrated that a long time span of moderately
accurate observations is more beneficial to accurate ephemerides of
natural satellites than a short period of extremely accurate data.

George Kaplan (USNO contractor) investigated the algorithms used for
predicting the relative positions of binary star components, given a
previously determined orbit.  He developed new computational code that
incorporates several geometric effects that have not previously been 
taken into account; these effects can be important for certain binary
systems, especially as observational accuracy improves.  He also
suggested that some conventions relating to how orbits are defined 
should be established for future work. 

Norbert Zacharias, on behalf of Erik H{\o}g gave a talk about absolute
astrometry in the next 50 years.  Erik H{\o}g proposes a 2nd Gaia-type
space mission to fly in roughly 20 years from now to significantly
improve the accuracy of proper motions and possibly go into the
near infra-red to better penetrate galactic dust.

This talk was followed by Anthony Brown's report on a recent meeting
regarding next steps towards future space astrometry missions, see also
camd08.ast.cam.ac.uk/ Greatwiki/GaiaScienceMeetings/FutureAstrometryJul15 .

Norbert Zacharias presented some evidence for astrophysical offsets 
between the radio and optical centers of emissions of compact, 
extragalactic sources (AGNs) which are envisioned as
link objects between the Gaia and ICRF reference frames.
At this point it is not clear if just the exclusion of some
sources with large structure will be sufficient or if a
general problem exists on the few mas level which would
render the accuracy of the radio-optical link much worse than
currently expected (Zacharias \& Zacharias 2014).

Nathan Secrest reported on identifying about 1.4 million sources
of the WISE catalog as very likely being AGNs.  This large pool
of sources will be valuable for QSO research in general and for
enhancing the Gaia to ICRF link of reference frames.

\section{Report 2012 -- 2015}

\subsection{Almanac and near Earth}

The 3rd edition of the Explanatory Supplement to the Astronomical
Almanac was published (Urban et al. 2012).
The US Naval Observatory's Nautical Almanac Office and Her Majesty's
Nautical Almanac Office (UK) updated all solar system computations,
incorporating JPL's DE430 ephemeris for the major bodies, using JPL's
Horizons ephemerides for minor planets and Ceres, and utilizing the value
of the astronomical unit consistent with the IAU 2012 resolution.
A paper explaining why the Meridian of Greenwich moved was published
(Malys et al. 2015).

A new idea for determining atmospheric refraction has been developed by 
utilizing differential measurements with double fields of view. 
In 2013 a prototype telescope with double fields of view was developed and
observations performed at very large zenith distances (Yong et al. 2015).

The OAFA (Astronomical Observatory Felix Aguilar) performed satellite 
observations. 
At the UNI (National Engineering University, Peru) the occultation of 
Mars by the Moon was observed at the AFARI Observatory. 

A collaboration between Tang Zheng Hong, Qi Zhao Xiang and Guo Sufen 
(Shanghai Observatory - China), A. Andrei, E. Nogueira and J.L. Penna 
(Nacional Observatory – RJ – Brazil) and R. Teixeira, M. Fiencio, 
M. Voelske and O. Rodrigues (IAG/USP – UNICSUL – Brasil) began an 
observational program of satellite debris at the Valinhos Observatory.
The Space Surveillance Telescope (SST) mainly used for space debris
detection also has some photometric and astrometric potential
(Monet et al. 2013).

From 2012 onward several initiatives combining the fields of Space Debris 
and Navigation Satellites have been started. At the Observatory of Tarija, 
Bolivia, a program of observation of space debris continued in collaboration 
with ISON (International Scientific Optical Network).  In Venezuela a 
combined program between the ABAE (Bolivarian Agency for Space Activities) 
and the CIDA (Astronomy Research Center) was started aiming to track 
geostationary satellites and space debris (Lacruz et al. 2015). 
In Brazil, fostered by the 
SHAO/CAS (Shanghai Astronomical Observatory of the Chinese Scientific 
Academy) 2 programs are running, one for space debris through ON (National 
Observatory) and USP (University of São Paulo), and  the other for navigation 
of satellites from the BEIDOU system at ON (National Observatory).
The Chilean telescope sites naturally harbor several such programs, as 
TAROT (French collaboration) and MODEST (University of Michigan 
collaboration). Several of these programs are associated with 
Time \& Frequency research, as for instance at the ONBA (Naval Observatory 
of Buenos Aires) and the IGNA (National Geographic Institute of Argentina),
in Argentina, and also at the ON (National Observatory) in Brazil.

\subsection{Solar System}

A set of astrometric solutions were developed (Qi \& Yu 2015a) of the 
pointing and tracking of celestial objects for the Lunar-based Ultraviolet 
Telescope (LUT), an astronomical instrument aboard Chang’e 3, the lunar probe 
of China’s Lunar Exploration Program that was successfully landed on the 
northern part of the Moon’s Mare Imbrium in late 2013. Feasibility was first 
shown with experiments done from Earth, and then confirmed with actual 
LUT observations from the Moon’s surface.

The Naval Observatory Flagstaff Station (NOFS) continued their long-term
program with the Flagstaff Automated Scanning Transit Telescope (FASTT)
monitoring Uranus, Neptune, Pluto and the bright satellites of 
Jupiter through Neptune for ephemerides improvement with about 300 
observations per year.
5700 bright asteroids are monitored with FASTT for ephemerides
improvement and for occultation predictions, with about 25,000 obs/year.
NOFS works with JPL to observe a few NASA targets with the 61-inch 
telescope for spacecraft navigation purposes.
Observations of Pluto and a sample of other bright Kuiper Belt Objects
were performed with the 61-inch telescope for occultation predictions.

2358 CCD Astrometric observations of the 5 major Uranian satellites were 
made between 1998 and 2007 near Beijing and Shanghai and compared with 
theory (Qiao et al. 2013) showing an accuracy of between 50 and 200 mas 
per observation.
Digitization and reduction of old astronomical plates of natural satellites 
began at the Chinese Academy of Sciences using a commercial scanner.

Triton's orbit was determined with the use of a revised pole model 
(Qiao, R.C. et al. 2014) using the 3108 Earth-based astrometric observations 
from the Natural Satellite Data Center (NSDC) covering the time span of 
1975 to 2006. A revised model for Neptune's pole was derived.

1095 astrometric positions of Triton were observed between 2007 and 2009 
(Zhang et al. 2014a) using 3 telescopes in China.  Comparison to ephemerides 
show residuals between 30 and 50 mas and significant differences between
various ephemerides.

Since 2011, the Jinan University group led by Q.~Peng has been
working on the geometric distortion (GD) solution for a CCD
frame and its application. This method was applied to CCD images
from the 1-m and 2.4-m telescope at Yunnan Observatory for open
clusters (Peng et al. 2012), the saturnian satellite Phoebe 
(Peng et al. 2015), and the near-Earth asteroid Apophis (Wang et 
al. 2015), improving the positional accuracy by about a factor of 2.

Differential VLBI was used to improve the ephemerides of Mars
and Saturn (Jones et al. 2015), and to track numerous spacecraft.
Most notably the MSL lander, Curiosity, in August 2012
was tracked with 1 nrad (200 $\mu$as) accuracy.

A remarkable success was obtained by the international group that includes 
several institutions of various South American and other countries for 
observation and interpretation of occultations when rings around the
asteroid Chariklo were observed at LaSilla, Chile by Brazilian astronomers.

\subsection{Solar Neighborhood}

The Research Consortium on Nearby Stars (RECONS) lead by T.~Henry, 
Georgia State Univ., continued their effort to obtain parallaxes and 
photometry of nearby stars.
About 150 M dwarfs could be added to the sample of stars within 25 pc
of the Sun (Finch et al. 2014, Riedel et al. 2014).

Trigonometric parallaxes and proper motions of young stellar objects 
and brown dwarfs were obtained by C.~Ducourant and J.F.~Le Campion 
(Bordeaux Observatory - France) and R.~Teixeira, P.A.B.~Galli, 
A.~Krone-Martins and A.C.~Ferreira (IAG/USP - Brazil),
see Galli et al.~2013, Ducourant et al.~2014, Teixeira et al.~2014. 
This research aims at determining membership and the study of properties
like ages and masses of stars as part of Fundamental Astronomy. 

A number of parallax programs were also concluded using the VLBA for
star-forming regions (Zhang et al. 2014b, Wu et al. 2014) and in the
infrared for brown dwarfs (Smart et al. 2013, Marocco et al. 2013).

NOFS continues the 61-in parallax observing program of red dwarfs
and white dwarfs which began about 20 years ago with papers in preparation.
Evaluation of the URAT data (see below) to obtain parallaxes of nearby
stars began.
NOFS has observed northern bright stars (magnitudes 0-6)
for a comparison with Gaia astrometry of bright targets.

\subsection{Double Stars}

Speckle astrometric studies of binaries have been carried out with the 
PISCO camera at Merate (Scardia et al. 2015) and lucky imaging of Luhman 
16AB (Mancini, et al. A\&A in press).
The result of Kaplan's investigation (see above) prompted a revision of
algorithms used in many orbital solutions e.g. the Washington Double Star 
program.

Work on the Washington Double Star catalog (WDS) continued which now 
contains 1,276,937 mean positions of 132,600 pairs. The USNO speckle
interferometry observing program primarily consists of observations 
obtained with the 0.67m refractor in Washington, DC (Mason et al. 2013,
Hartkopf et al. 2015).
Over the triennium 7367 observations were 
made resulting in 3607 mean positions. Collaborations continue with 
Tokovinin (CTIO), Roberts (JPL), and others to obtain measures of 
pairs inaccessible with the USNO refractor (Tokovinin et al. 2015).
For details on the double cataloging and observing programs of the 
USNO please see the Commission 26 Triennial Report.

The CHARA optical interferometer can perform astrometric observations
of binary systems with high angular resolution (ten Brummelaar, Tuthill
\& Van Belle, 2013, Farrington et al 2014).  
It continues to work on spectroscopic binary stars, Be Stars, separate
fringe packet objects and the detection and imaging of faint companions
to many classes of objects.
Gail Schaefer at CHARA, continues a binary star orbit program using the 
Adaptive Optics system at the Keck Observatory.

\subsection{Star Catalogs}

The final release of the Carlsberg Meridian Catalogue series (CMC15) was
published (Muinos \& Evans 2014).  It is the last of the series and
comprises all the observations made between March 1999 and 
March 2011 with the Carlsberg Automatic Meridian Circle in El Roque de los
Muchachos Observatory on the island of La Palma (Spain). The catalogues
CMC12, CMC13, and CMC14 are superseded by this one. CMC15 contains more
than 122 million observations in the magnitude range of 9 $\le r' \le$ 17 
and declination range of $-40^{\circ} \le \delta \le +50^{\circ}$. 
The catalogue internal 
errors in astrometry are below 30 mas in both coordinates for stars brighter
than $r'$=13, reaching 60 mas for $r'$=16. The internal magnitude error is 
below 0.020 mag for stars brighter than $r'$=13, and about 0.090 mag for 
$r'$= 16.  The instrument has been decommissioned and will go into a museum.
Work continued on the SST (Space Surveillance and Tracking) field with 
the Telescope Fabra-ROA at Montsec (TFRM).

The USNO CCD Astrograph Catalog project concluded with the release of
the final, 4th catalog, the UCAC4 (Zacharias et al. 2013).  This all-sky
astrometric survey covers stars between about R = 7 and 16 with 20 to 70 
mas positional accuracy depending on magnitude.

A complete re-design of the USNO ``redlens" astrograph (which also was used
for the UCAC project) utilizes a 28 sq.deg. focal plane consisting
of 4 STA1600 CCDs with 10,560 by 10,560 pixels each.  The resulting USNO 
Robotic Astrometric Telescope (URAT) project operated for 3 years 
(April 2012 to June 2015) from NOFS.  A first data release,
the URAT1 catalog (Zacharias et al. 2015) provides accurate positions
of over 220 million stars between about Dec = +90 and $-$15 deg and about
R = 4 to 18.5 mag.  The large number of observations per star (average 24)
results in small random errors with an estimated systematic error floor
of about 10 mas.  The URAT data also allow to derive parallaxes
of all nearby stars in that area of sky, the largest survey of this
kind since the Hipparcos program.  URAT was relocated to CTIO in
mid 2015 and began a southern hemisphere survey aiming mainly at
the brightest stars where Gaia's capabilities are not well understood
at this time.

The UNC (National University of Cordoba, Argentina) established a catalog 
of the ecliptic area.
At the CIDA (Astronomy Research Center, Venezuela), a combined 
astrometric catalogue is being compiled using Carte du Ciel and 
Carlsberg-CAMC fields.

Several investigations and comparisons of large astrometric catalogs of 
positions and proper motions were performed (Vityazev \& Tsvetkov 2014).
An updated version, the XPM2 catalog was constructed (Fedorov, Akhmetov, \&
Shulga 2014) based on SuperCOSMOS data of Schmidt plate scans and linked 
to extragalactic sources.
Vector spherical harmonics algorithms were developed for astrometric
catalog analysis (Mignard \& Klioner 2012).

A new catalog of absolute proper motions and updated positions was 
presented (Qi \& Yu 2015b), derived from the Space Telescope Science 
Institute digitized Schmidt survey plates utilized for the
construction of Guide Star Catalog II. As special attention was devoted 
to high accuracy by the removal of position, magnitude, and color dependent 
systematic errors through the use of both stars and galaxies, this release 
is solely based on plate data outside the galactic plane 
$|b| \ge 27^{\circ}$.

Absolute astrometry was obtained from PanSTARRS data in collaboration 
with USNO (Makarov et al. 2015).  The accuracy of the obtained proper 
motions matches the simulations while the observed positional accuracy 
of pre-public release data is significantly worse than expected. 
A major public release of PanSTARRS data is expected by end of 2015.
Future astrometry with LSST was further investigated (Ivezic et al. 2013).
The Palomar Transient Factory 2nd data release became public in August
2015 (http://www.ptf.caltech.edu/iptf). Its image processing and data
archive is explained in Laher et al. 2014.

\subsection{Proper Motions and our Milky Way Galaxy}

NOFS completed a deep proper motion survey in the SDSS imaging area
(Munn et al. 2014) with a followup paper in preparation on the white 
dwarfs in the survey.

Various projects for the determination of proper motions from ground-based,
wide-field imagers (Libralato et al. 2014, 2015) the Hubble Space Telescope 
(Bedin et al. 2014, Massari et al. 2015) and photographic plates 
(Qi \& Yu 2015b) have been concluded.
The kinematics of our Milky Way X-shaped bulge was analyzed using proper
motion observations (V{\'a}squez et al. 2013).

The absolute proper motions of 3 globular clusters (NGC 6397, 6626 and
6656) were measured based on Southern Proper Motion (SPM) material 
(Casetti-Dinescu et al. 2013).
Extensive use of the SPM4 catalog was made to select OB-type candidate stars 
and then study these spectroscopically. Recent results indicate the 
existence of such stars at the edge of the Galactic disk, likely formed 
recently due to the interaction between the Magellanic Clouds and the 
disk of the Milky Way (Casetti-Denescu et al. 2012).
SPM4 proper motions were used to study some 8000 RR Lyrae stars from the 
recent Siding Springs catalog. An RR Lyrae overdensity was found towards 
the inner regions of the Milky Way that is kinematically distinct from the 
Milky Way stellar populations (Casetti-Dinescu et al. 2015). 
Proper motions indicate the overdensity has a net vertical motion away 
from the Galactic plane. The possibility that this overdensity is debris 
from Omega Centauri's parent galaxy was explored but no definite conclusion
can be drawn at this time.

Possible systematic errors in the Sloan Digital Sky Survey (SDSS) proper 
motions were investigated by comparing them with proper motions from the
Kapteyn Selected Area (SA) survey in order to estimate the accuracy of the 
absolute proper motion zero point. Results from 22 SA fields indicate that
the SDSS zero point has an uncertainty of about 1 mas/yr in fields with 
low reddening, while in fields with large reddening errors are of a 
few mas/yr (Ahn et al. 2012).

The Visible and Infrared Survey Telescope (VISTA) was used to study
high proper motion objects in the galactic plane (Smith et al. 2014,
Kurtev et al. 2015).
The Hubble Space Telescope was used for a dedicated proper motion
project (HSTPROMO) reaching globular clusters (Bellini et al. 2014,
Watkins et al. 2015)  and members of the local group of galaxies
(Platais et al. 2015).  
The proper motion of the Large Magellanic Cloud was also obtained from
ground-based observations using the SPM catalog (Vieira et al. 2014).

\subsection{International Celestial Reference Frame (ICRF)}

The journal paper about the second realization of the International
Celestial Reference Frame (ICRF2) by Very Long Baseline Interferometry
(VLBI) was published (Fey et al. 2015).

The ICRF3 working group was reformed in 2012 after a 3-year hiatus 
following ICRF2.  The 2012-2015 triennium focussed on assessing the 
needs and initiating observations to address those needs, aiming at
adoption of the ICRF3 by the IAU GA in 2018.
VLBA S/X Cal Survey II (VCS-II) was carried out under David Gordon's 
leadership.  About 2400 sources were detected and precision improved 
almost by a factor of 4. The K-band CRF became full-sky and is being 
densified under the leadership of PI Alessandra Bertarini.
The X/Ka-band became full-sky in 2012. In 2015 its precision surpassed 
that of the ICRF2.
In late 2012 the X/Ka CRF work expanded to incorporate a collaboration
with ESA and their 35-m antenna in Malargue Argentina.
Observational programs were put in place to increase the number
of radio sources with optically bright counterparts (V $\le$ 18 mag)
for an improved Gaia to ICRF reference frame tie.
Observations are now underway at S/X, K, and X/Ka-bands.

A new combined catalog of radio source positions was computed
(Sokolova \& Malkin, 2014).
Impact of correlations between radio source coordinates in VLBI-based 
catalogs on the orientation angles between two CRF realizations was 
estimated (Sokolova \& Malkin, 2012).
Apparent and proper motions of ICRF radio sources were further studied
(Voronkov \& Zharov 2013, Zharov et al. 2014).

The affect of Galactic aberration on the ICRS realization and on the Earth
orientation parameters (EOP), which refer to the ICRS were investigated
(Liu et al. 2012, Liu, Xie \& Zhu 2013). Although the effect is small 
(few microarcseconds) it is being considered by several IAU working 
groups of Division A.
Impact of the Galactic aberration on proper motions of the celestial 
reference frame was further investigated 
(Liu et al. 2012, Malkin 2012, 2014, 2015).

The ICRS product center of the IERS evaluates the consistency of celestial 
reference frames produced at different IVS analysis centers by re-analysis 
of full VLBI observational database on a yearly basis. This evaluation is 
done through the modelling of the coordinate difference between the ICRF2 
defining source coordinates in both the individual catalogs and the ICRF2.
Results are shown in the successive annual reports of the IERS (see 
http://www.iers.org/IERS/EN/
Organization/ProductCentres/ICRSCentre/icrs.html).

\subsection{Radio-Optical Reference Frame Link}

Offsets between the centers of emission as observed in the radio and 
optical were found for ICRF sources (Zacharias \& Zacharias 2014),
see also above (section GA business meeting).
Staff from the Paris Observatory, as part of the ICRS product Center 
of the IERS, were deeply involved in the monitoring of the ICRF sources 
observed in VLBI, the study of their optical counterparts and the 
compilation of all the recorded quasar names in the Large Quasar 
Astrometric Catalogue (LQAC).

An observing campaign is under way using the T-1m class telescopes with
automation (TAROT). They are observing regularly a selected sample of 
AGN’s well suited for the radio-optical link between the ICRF2 and the 
future Gaia catalogue. 
In particular studies of correlations between astrometric positions and 
flux variations as well as very regular photometric time series have 
been undertaken (Taris et al. 2011, 2013). Statistical analysis of 
these series will be performed to determine periodic signatures due to 
astrophysical processes.

Due to the drastic increase of the number of discovered quasars, mainly 
from the SDSS catalog updates, 2 new releases of the LQAC, respectively 
the LQAC2 (Souchay et al. 2012) and the LQAC3 (Souchay et al. 2015) were 
delivered, the last one including 321,957 objects. In addition to 
giving the information obtained from the original catalogues as 
redshift and multi-bandpass magnitudes, the LQAC contains new or 
improved data from re-calculations of quasar coordinates, 
estimated absolute magnitudes and determination of morphology indexes.
Moreover some statistical and physical studies have been performed
using the LQAC2 catalogue, including an estimate of about 1 million for
the number of quasars expected to be detected by the Gaia mission 
(Gattano et al. 2014).

\subsection{VERA extragalactic Radio Astrometry}

The Mizusawa VLBI Observatory of NAOJ continued the operation 
of VERA (VLBI Exploration of Radio Astrometry), which is a dedicated array 
for maser astrometry. To date, parallaxes and proper motions have been 
obtained for 40 Galactic maser sources, namely, star-forming regions or 
late-type stars. Recent results of VERA have been published in the VERA 
special issue of PASJ in 2014 Vol. 66-6 and 2015 Vol. 67-4. 
Highlights are the discussion of systematic deviations of source motions 
from the Galactic rotation in terms of the density-wave spiral-arm model
(Sakai et al. 2015), and the calibration of period-luminosity relation of 
Miras based on accurate distances measured with VLBI (Nakagawa et al. 2014).
The determination of Galactic fundamental constants is reported
(Honma et al. 2012) based on 52 star-forming regions for which accurate 
astrometric results were available, showing that R0 = 8.05 $\pm$ 0.45 kpc 
and Theta\_0 = 238 $\pm$ 14 km/s. 
By combining the results of VERA, VLBA (Very Long Baseline 
Array) and EVN (European VLBI Network), VLBI maser astrometry has been 
obtained for more than 100 sources (Reid \& Honma 2014).

The Japanese astronomical community established the Japan Square Kilometre 
Array Consortium (SKA-JP) in 2008, then the SKA-JP Astrometry Working 
Group in 2009. This WG activated real astrometry demonstrations at 
the low frequency band (1.6 GHz) since 2013. The first successful 
detection of a trigonometric parallax was announced in the autumn 
meeting of the Astronomical Society of Japan (H. Imai and the SKA-JP 
Astrometry Working Group, 2015 September). This result suggests the 
high potential of high precision radio astrometry (10 micro-arcsecond 
level) at the low frequency bands covered with the SKA (SKA1-MID Band-2,
SKA1 System Baseline Design, 2013 March). The WG has provided a core 
member of the SKA VLBI Working Group (established in 2015 August) for 
realizing the SKA specification available to high precision radio 
astrometry with the SKA. The WG presented its idea of future 
radio astrometry with the SKA at the 1.6 GHz band for exploring the 
dynamical structure of the Milky Way and the Local Group 
(e.g. Imai et al., in the Asia-Pacific Radio Science Conference 2013).

The KVN and VERA array (KaVA) has been partly opened as an open-use array 
toward Japanese, Korean and Taiwanese communities since 2013. Compared 
to the VERA array, KaVA has an advantage for observations of faint and 
extended sources. Regarding the astrometry capability of KaVA, 
evaluations and test observations are currently conducted to open 
the KaVA astrometry mode by March 2016 or later.

\subsection{Other Reference Frame Topics}

The direct distance determination process using trigonometric
parallaxes was reviewed in light of physical assumptions
(Lindegren 2013).

The multi-wavelength catalogues of 2MASS, WISE, and AKARI are used to 
calculate the position of the Galactic plane (Ding, Zhu \& Liu 2015).
The parameters for the direction of the North Galactic pole and the 
Galactic center are given which can be used to redefine the Galactic 
coordinate system by IAU in the future.
The definition and use of the ecliptic in modern astronomy was
reviewed (Capitaine \& Soffel 2015).

After more than 10 years since its adoption by the IAU in 2006 the
precession model was studied again (Liu \& Capitaine 2015) using new 
solutions of the Earth-Moon Barycenter (EMB) motion, new theoretical 
contribution to the precession rates, and a revised J2 long-term 
variation obtained from Satellite Laser Ranging (SLR).

\subsection{JASMINE}

JASMINE is an acronym for Japan Astrometry Satellite Mission for 
Infrared Exploration. Three satellites are planned as a series of JASMINE 
projects, as a step-by-step approach, to overcome technical issues and 
promote scientific results (Gouda 2011, Gouda 2012). These are Nano-JASMINE,
Small-JASMINE and (medium-sized) JASMINE. 
Nano-JASMINE is a nano-size satellite of 50 cm size that weights 
about 35 kg (Hatsutori et al. 2011). 
The diameter of the primary mirror is 5 cm. A fully depleted CCD is located 
at the focal plane of the telescope (Kobayashi et al. 2010). 
The flight model of the Nano-JASMINE satellite was fabricated in Oct. 2010.
Nano-JASMINE will operate in the zw-band (0.6 to 1.0 micron). The target 
accuracy of parallaxes is about 3 mas at zw=7.5 mag (Kobayashi et al. 2011).
Experimental evaluation of radiation damage has been performed
(Kobayashi et al. 2012).  Moreover high-accuracy proper motions 
(0.1 mas/year) can be obtained by combining the Nano-JASMINE 
catalogue with the Hipparcos catalogue (Michalik et al. 2012).  
The observing strategy and methods used in the data analysis for Nano-JASMINE
will be similar to what is planned for Gaia. Hence the use of Nano-JASMINE 
data is useful to check algorithms that are to be used in the Gaia data 
analysis, and the algorithms of Gaia data analysis can be applied 
to Nano-JASMINE (Yamada et al. 2012).
Economic and political issues prevented an early launch by a Ukraine-Brazil
launch service.  Nano-JASMINE is expected to be launched into a
Sun-synchronized orbit with an altitude of about 650 to 800 km in late 2017.

Small-JASMINE will determine positions and parallaxes accurate to about 
10 to 20 $\mu$as for stars towards a region around the Galactic nuclear 
bulge and other small regions that include scientifically interesting 
target stars (e.g. Cyg X-1), brighter than Hw=11.5 mag (1.1 to 1.7 
micron).  Thus it will be complementary to Gaia.
Proper motions of between 10 and 50 $\mu$as/year are 
expected. The survey will be done with a single beam telescope of which 
the diameter of the primary mirror is about 30 cm (Yano et al. 2011). 
The target launch date is around 2021. The basic designs of Small-JASMINE 
and technical problems have been investigated mainly at the JASMINE project 
office of the National Astronomical Observatory of Japan in collaboration 
with JAXA, Kyoto University and other institutes (Utsunomiya et al. 2014).

The main scientific objective of Small-JASMINE is to clarify the dynamical 
structure of the Galactic nuclear bulge and search for observational 
relics of a sequential merger of multiple black holes to form the 
supermassive black hole at the Galactic center. In particular, 
Small-JASMINE's primary purpose well be to provide an understanding of 
the past evolution process of the supermassive black hole and the prediction 
of future activities of our Galactic center using the knowledge of the 
gravitational potential in the Galactic nuclear bulge. This understanding 
can contribute to a better understanding of the co-evolution 
of the supermassive black holes and bulges in external galaxies.

(Medium-sized) JASMINE is an extended mission of Small-JASMINE, which 
will observe towards almost the whole region of the Galactic bulge with 
accuracies of 10 $\mu$as in Kw-band (1.5 to 2.5 micron). The target 
launch date is the 2030s.  Merging with other future missions like WFIRST
or post-Gaia ESA mission is under consideration.

\subsection{Gaia}

The ESA Gaia mission was launched on 19 December 2013, and after a 6-month
commissioning period started its routine science observations in July 2014.
The main scientific aim of Gaia is to reveal the structure and kinematics 
of our Galaxy. The science requirements deduced from the main goal and
amended by many other science cases have resulted in a mission conducting 
an astrometric, photometric and spectroscopic survey of the full sky. 
Gaia is anticipated to detect and measure more than 1 billion objects 
astrometrically (Lindegren et al. 2012) and photometrically. 
In addition, spectroscopy will provide radial velocities for an 
estimated 150 million stars.

During the commissioning period a number of anomalies were encountered, 
in particular excess stray light entering the payload, the much larger 
than expected variations of the basic angle between the lines of sight 
of Gaia's two telescopes, and transmission degradation due to continued 
out-gassing (http://www.cosmos.esa.int/web/gaia/news\_20140729). 
On-board measures were implemented to mitigate the effect of the stray 
light for in particular the Radial Velocity Spectrograph (RVS). The
effect of the additional stray light has been accounted for in updated 
performance predictions for the Gaia mission 
(http://www.cosmos.esa.int/web/gaia/science-performance). 
The transmission loss was countered by decontamination activities, the 
last of which took place in June 2015 (the previous one in September 2014).
The contamination rate appears to be slowing. The basic angle variations 
are being measured to high precision with the on-board metrology system and
investigations are ongoing to account for the variations in the astrometric 
processing through a combination of metrology measurements and 
self-calibration from the data (Lammers \& Lindegren 2014).

Over the first year of science operations a number of results were 
presented that highlight the capabilities of Gaia (Brown 2015).
These include the 
first supernova discovery, the showcasing of spectra obtained with the 
RVS and of spectrophotometry obtained with the prism photometers, 
demonstrations of the detection and observation of solar system objects, 
a demonstration of the excellent quality of the on-board detection and 
initial data treatment through observations of Einstein's cross, and
two examples of how Gaia deals with crowded regions and nebulae seen in 
emission. At the IAU General Assembly in Honolulu a first Gaia 
Hertzsprung-Russell diagram, based on the so-called Tycho-Gaia 
Astrometric Solution, was presented. 
It demonstrates the overall correctness and readiness of the data
processing chain, the quality of the Gaia instruments, and strengthens
the confidence that the impact of the Basic Angle variations can be 
eliminated from the final astrometric results.

The Gaia Data Processing and Analysis Consortium (DPAC) swung into action 
straight after the launch and provided support to the commissioning 
activities through its Initial Data Treatment (IDT) and First Look (FL) 
pipelines as well as through the analysis of data by the DPAC payload 
experts.  During the first year of science operations the main active 
DPAC systems were the IDT/FL pipelines, the Astrometric Verification 
Unit's basic angle monitor and astrometric instrument model pipelines,
the RVS processing pipelines provided by the DPAC payload experts, 
the photometric processing pipeline, and the Astrometric Global Iterative 
Solution system, responsible for the core astrometric data processing. 
In parallel a number of the advanced processing pipelines (variable stars,
astrophysical parameters, solar system objects, non-single stars) were 
exercised on the data collected over the special Ecliptic Pole scanning 
period during the first month of science observations. Throughout the 
period since launch the ground based optical tracking of Gaia took
place through regular observations of the spacecraft from the 2.0-m 
Liverpool Telescope on La Palma, ESO's 2.6-m VST on Paranal, and the 
Las Cumbres telescopes. These observations will be included in
the very demanding orbit determination for the Gaia spacecraft.

At the time of writing this report ESA and DPAC are gearing up toward 
the first Gaia data release in 2016.

\subsection{Gaia related}

Gaia data will allow to test aspects of fundamental physics
like gravitational deflection of light, energy flux of gravitational
waves and Local Lorenz Invariance (Klioner 2014).
The astrometric exoplanet detection capability with Gaia was investigated
(Perryman et al. 2014).

Many astronomers in Europe, including but by far not limited to 
astrometrists, are involved in Gaia supporting activities,
particularly the data processing centers.
For more details see the ESA Gaia web pages.

\subsection{Other Space Missions}

The following astrometry related space missions are in the planning stages:
the gravitation astrometric measurement experiment (Gai et al. 2012a), 
interferometric stratospheric astrometry for solar system objects 
(Gai et al. 2012b) and the production of the Fine Guidance Star Catalog for
Euclid in collaboration with Thales-Alenia (Bucciarelli et al. 2015).

Erik H{\o}g is proposing a 2nd Gaia-type mission to significantly improve
the proper motions (H{\o}g 2015).  Ideally such a mission would fly about
20 years after Gaia and moving the bandpass into the IR is under
consideration.

\subsection{Teaching}

A collaboration between R. Teixeira, J.P. Delicato, D. Miranda and 
M. Fidencio (IAG/USP – Brazil) started a teaching program about the use 
of the concept of space in science education, science fiction setting in 
teaching Astronomy and activity book for remote observations with 
educational purposes.

\subsection{Meetings}

The Journ{\'e}es meeting scientific developments from highly
accurate space-time reference systems, was held at Paris Observatory
in September 2013.
The IAU-sponsored conference Journ{\'e}es 2014, organized in cooperation 
with the Paris Observatory was held in Pulkovo Observatory in September 2014.
The 2 most recent  ADeLA (Dynamical Astronomy in Latin America) meetings
took place in Argentina (2012, 5th meeting) and Chile (2014, 6th meeting).
Volume 46 of RMxAA is dedicated to the proceedings of ADeLA 6.
IAU Symposium 298 (Setting the scene for Gaia and LAMOST) was held
at Lijang, China in May 2013.
Proceedings of the IAU Symposium 289 (Advancing the Physics of Cosmic
Distances) was published in 2013.

\vspace*{3mm}
\normalsize
\noindent
{\bf References}

%\small

\begin{flushleft}
Ahn, C.P. et al. 2012, ApJS 203, 21 (SDSS DR9)

Bedin, L. R., et al., 2014 MNRAS 439 (proper motions, HST)

Bellini, A. et al. 2014, ApJ 797, 115 (HSTPROMO globular clusters)

Brown, A. 2015, IAU GA meeting 29, no. 2252804 (Gaia data processing)

Bucciarelli et al. 2015 AASP 42 (FGS for Euclid)

Capitaine, N. \& Soffel, M. 2015, Proc. Journ{\'e}es 2014, p.61-64 
  (ecliptic)

Casetti-Dinescu, D.I. et al. 2012 ApJ 753, 123 (OB star kinematics)

Casetti-Dinescu, D.I. et al. 2013 AJ 146, 33 (space velocities glob.clusters)

Casetti-Dinescu, D.I. et al. 2015, ApJ 810, L4, (RR-Lyra overdensity)

Ding, P.-J., Zhu, Z., \& Liu, J.-C. 2015, Research in A \& A.
   15, 1045 (galactic plane)

Farrington, C. et al. 2014, AJ 148, 48 (CHARA)

Fedorov, P.N., Akhmetov, V.S., Shulga, V.M. 2014, MNRAS 440, 624 (XPM2)

Fey, A. et al. 2015, AJ, 150, 58 (ICRF2)

Finch, C. et al. 2014, AJ 148, 119 (UCAC4 nearby stars)

Gai, M., et al. 2012a ExA 34 (gravitation astrometric mission) 

Gai, M., et. al. 2012b SPIE 8446 (interferom. stratosph. astrometry)

Galli, P.A.B. et al. 2013, A\&A 558, A77 (proper motions moving cluster)

Gattano,C., Souchay,J., Barache,C. 2014, A\&A, 564A, 117G (LQAC-2)

Gouda, N. 2011, Scholarpedia, 6(10),12021 (JASMINE)

Gouda, N. 2012, ASP Conf. Proc., 458. San Francisco, CA: p.~417 (JASMINE)

Hartkopf, W.I. et al. 2015, AJ 150, 136 (USNO speckle interferometry)

Hatsutori, Yoichi et al. (2011). ESA publ. 45, 397 (Gaia)

H{\o}g, E. 2015, https://dl.dropbox.com/u/49240691/GaiaRef.pdf  (Gaia2) 

Honma M. et al. 2012, PASJ, 64, 136 (VERA, galactic constants)

Ivezic, Z. et al. 2013, AAS meeting 221, abstr.247.06 (LSST astrometry)

Jones et al. 2015 AJ (VLBI ephemerides Mars, Saturn)

Klioner, S.A. 2014, EAS Publ.Series 67-68, 49-55 (Gaia, fundamental physics)

Kurtev et al. 2015 RMxAA Serie Conferencias 46, 50 (VISTA proper motions)

Kobayashi, Y. et al. 2010, Proc. SPIE, 7731, 77313Z-1 (Nano-JASMINE)

Kobayashi, Y. et al. 2011, ESA publ. 45, 401 (Nano-JASMINE)

Kobayashi, Y. et al. 2012, Proc. SPIE, 8442, 844247, 6 (Nano-JASMINE)

Lacruz et al. 2015, RMxAA Serie Conferencias 46, 87 (geo-sat., space debris)

Laher, R.R. et al. 2014, PASP 126, 674 (Palomar Transient Factory)

Lammers, U. \& Lindegren, L. 2014, ASP Conf.Series 485, 491 (Gaia calibration)

Libralato, M. et al. 2014, A\&A 563 (proper motions, wide-field)

Libralato, M. et al. 2015, MNRAS 450 (proper motions, wide-field)

Lindegren, L. 2012, A\&A 538, 78 (astrometric core solution Gaia)

Lindegren, L. 2014, in Proc. IAU Symp. 289, p.52 (trig. parallaxes)

Liu, J.-C. et al. 2012, A\&A, 548, A50 (galactic aberration)

Liu, J.-C., Xie, Y., \& Zhu, Z. 2013, MNRAS, 433, 3597 (galactic aberration)

Liu, J.-C. \& Capitaine, N. 2015, in Proc. Jounr{\'e}es 2014, 
    155-158 (precession model)

Makarov, V. et al. 2015, IAU GA 225, 7836 (absolute astrometry PanSTARRS)

Malkin Z. 2012, Proc. Journ{\'e}es 2011 Vienna, Austria, p.168-169
 (galactic aberration)

Malkin Z. 2014, MNRAS, 445(1), 845-849 (galactic aberration)

Malkin Z. 2015, MNRAS, 447(4), 4028 (galactic aberration)

Malys, S. et al. 2015, J.of Geodesy DOI 10.1007/s00190-015-0844-y
  (Greenwich)

Marocco, F., et al., 2013 AJ 146 (parallaxes brown dwarfs)

Mason, B., Hartkopf, W. \& Hurowitz, H. 2013, AJ 146, 56 (USNO speckle interf.)

Massari, D., et al., 2015 ApJ 810 (proper motions, HST)

Melis, C. et al. 2014, Science 345, 1029 (VLBI Pleiades distance)

Michalik, Daniel et al. (2012) ASP Conf. Series, 461, 549 (JASMINE - Hipparcos)

Mignard, F. \& Klioner, S. 2012, A\&A 547, A59 (spherical harmonics)

Monet, D. et al. 2013, AAS meet. 221, abstr.~352.17 (SST)

Muinos, J.L., Evans, D.W. 2014, AN 335, n.4, p.367 (CMC15)

Munn et al. 2014, AJ 148, 132 (proper motions in SDSS area)

Nakagawa A. et al. (2014), PASJ, 66, 101 (VERA Miras period-luminosity)

Peng, Q. et al. 2012 AJ 144, 170 (CCD distortion corrections)

Peng, Q. et al. 2015 MNRAS 449, 2638 (saturnian satellite Phoebe)

Perryman, M. et al. 2014, ApJ 797, 14 (Gaia exoplanet detection)

Platais, I. et al. 2015 AJ 150, 89 (HST proper motion LMC)

Qi Z. \& Yu Y. 2015a, PASP Nov. (Lunar Telescope)

Qi Z. \& Yu Y. 2015b, AJ, (DOI 10.1088/0004-6256/150/4/137)
 (photogr.plates PM)

Qiao, R.C. et al. 2013, MNRAS, 428, 2755 (Uranian satellites)

Qiao, R.C. et al. 2014, MNRAS, 440, 3749 (Triton orbit)

Reid M. J. \& Honma M. (2014), ARA\&A, 52, 339 (VERA, VLBI, VLBA astrometry)

Riedel, A.R. et al. 2014, AJ 147, 85 (solar neighborhood XXXIII)

Sakai N. et al. 2015, PASJ, 67, 69 (VERA, galactic rotation)

Scardia, M., et al. 2015 AN 336 (speckle, double stars)

Smart, R. L., et al., 2013 MNRAS 433 (parallaxes brown dwarfs)

Smith, L. et al. 2014, MNRAS 443, 2327 (VISTA proper motions gal.plane)

Sokolova Ju., Malkin Z. 2012, In: IVS 2012 General Meeting Proc., Eds. \\
  \hspace*{5mm} D. Behrend, K.D. Baver, NASA/CP-2012-217504, 339-341
 (CRF compar.)

Sokolova Yu.R., Malkin Z.M. 2014, Astronomy Letters, 40, 5, 268-277 
  (radio cat.) 

Souchay, J et al. 2012, A\&A 537A,99S  (LQAC2)

Souchay, J. et al. 2015, A\&A in press (LQAC3)

Taris, F. et al. 2011, A\&A 526A 25 (CFHTLS Deep-2 QSO’s)

Taris, F. et al. 2013, A\&A, 552A, 8 (astrom. photom. corr. opt. ICRF)

ten Brummelaar,T., Tuthill,P. \& Van Belle, G. 2013, J. of Astron.
  Instr. 2 (CHARA)

Tokovinin, A. et al. 2015, AJ 150, 50 (speckle interferom. SOAR)

Urban, S.E. et al. 2012, 
  3rd Ed. of Explanatory Supplement to the Astronomical \\
  \hspace*{5mm} Almanac, Univ. Science Books, Mill Valley, CA

Utsunomiya, Shin et al.(2014) Proc. of the SPIE, 9143, 91430Z 6
  (small-JASMINE)

V{\'a}squez, S. et al. 2013 A\&A 555, A91 (x-shaped galactic bulge)

Vieira, K. et al. 2014, RMxAC 43, 51 (proper motion LMC, SPM)

Vityazev V.V., Tsvetkov A.S. 2014, MNRAS, 442, p.1249-1264 (catalog comp.)

Voronkov N.A., \& Zharov V.E. 2013,
  Moscow Univ. Physics Bull., 68, 3, 235-240 \\
  \hspace*{5mm} (apparent proper motions ICRF)

Wang, N. et al. 2015 MNRAS 454, 3805 (near Earth asteroid)

Watkins, L. et al. 2015, ApJ 812, 149 (HSTPROMO globular clusters)

Wu, Y. W., et al., 2014 A\&A 566 (parallaxes VLBA)

Yamada, Y. et al. 2012, ASP Conf.~Series, 461, 585 (Nano-JASMINE, Gaia)

Yano, T. et al. 2011, ESA publ. 45, 449 (small-JASMINE)

Yong, Y. et al. 2015, Res. in Astron. \& Astroph. 15(10), 1742-1750 \\
  \hspace*{5mm} (atmospheric refraction)

Zacharias, N. et al. 2013, AJ 145, 44 (UCAC4)

Zacharias, N. et al. 2015, AJ 150, 101 (URAT1)

Zacharias, N. \& Zacharias, M.I. 2014, AJ 147, 95 (radio-optical link)

Zhang,H.Y. et al. 2014a, MNRAS, 2014, 438, 1663 (Triton positions)

Zhang, B., et al., 2014b ApJ 781 (parallaxes VLBA)

Zharov V. et al. 2014, Proc. Journ{\'e}es 2013, p. 73-76 
  (apparent proper motions ICRF)
\end{flushleft}

\end{document}